\documentclass[conference]{IEEEtran}
\IEEEoverridecommandlockouts
\usepackage{cite}
\usepackage{amsmath,amssymb,amsfonts}
\usepackage{algorithmic}
\usepackage{graphicx}
\usepackage{textcomp}
\usepackage{xcolor}
\usepackage{multirow}
\usepackage{array}
\usepackage{longtable}
\usepackage{hyperref}
\usepackage{framed}
\usepackage[export]{adjustbox}
\def\BibTeX{{\rm B\kern-.05em{\sc i\kern-.025em b}\kern-.08em
    T\kern-.1667em\lower.7ex\hbox{E}\kern-.125emX}}
\begin{document}

%\title{Resounding the Foundations: Revisiting Physics of Notations for Modelling with Symbolic Sound}

\title{Unleashing the Power of Sound: Revisiting the Physics of Notations for Modelling with auditory symbols}

%\author{}
\author{
\IEEEauthorblockN{Nuno Guerreiro}
\IEEEauthorblockA{\textit{NOVA LINCS, FCT NOVA University} \\
Lisbon, Portugal \\
n.guerreiro@campus.fct.unl.pt}
\and
\IEEEauthorblockN{Vasco Amaral}
\IEEEauthorblockA{\textit{NOVA LINCS, FCT NOVA University} \\
Lisbon, Portugal \\
0000-0003-3791-5151}
\and
\IEEEauthorblockN{Miguel Goulão}
\IEEEauthorblockA{\textit{NOVA LINCS, FCT NOVA University} \\
Lisbon, Portugal \\
000-0002-5356-5203}
}

\maketitle

\begin{abstract}
Sound - the oft-neglected sense for Software Engineering - is a crucial component of our daily lives, playing a vital role in how we interact with the world around us. In this paper, we challenge the traditional boundaries of Software Engineering by proposing a new approach based on sound design for using sound in modelling tools that is on par with visual design. By drawing upon the seminal work of Moody on the `Physics' of Notations for visual design, we develop a comprehensive catalogue of principles that can guide the design of sound notations.

Using these principles, we develop a catalogue of sounds for UML and report on an empirical study that supports their usefulness. Our study lays the foundation for building more sophisticated sound-based notations. The guidelines for designing symbolic sounds for software models are an essential starting point for a new research thread that could significantly and effectively enable the use of sound in modelling tools.

%We then propose a catalogue of sounds for UML
% based on the collected design principles and report on an empirical study to support their usefulness. This study lays the foundations for building more sophisticated and effective sound-based notations by providing a standardised set of design principles. The guidelines for designing symbolic sounds for software models are an essential starting point for a new research thread that could significantly and effectively enable the use of sound in modelling tools.
\end{abstract}


\section{Introduction}
Visual notations are well accepted in the Computer Science field \cite{kuhar2021conceptualization}. However, the tools that accommodate this diagrammatic communication do not support sound in the context of said notations. This relative lack of interest regarding sound potentially leads to the exclusion of many visually impaired or blind people from software development-related activities, namely software modelling with visual notations. One such example is UML. Created to standardise notational systems in software design, UML is a general-purpose, widely adopted modelling language and, as such, an integral part of the software development process \cite{brambilia2012MDSE}.

We argue that sound should be treated in the context of Software Engineering as it is already done with the visual approaches. It can potentially contribute to modelling tools, significantly increasing productivity and accessibility.

To assemble an effective interface and assign meaning to sound, we need to understand what influence different acoustic cues have through their cataloguing and systematisation. %It is shown that organised 
Organised sounds, such as music, play a role in memory creation and recollection, and auditory cues influence human behaviour \cite{blake2015acoustic}. This is relevant to our work, as we intend to achieve a user-friendly interface that does not require much effort from its users. As noted by Nielsen, humans have limited short-term memories~\cite{nielsen1994heuristics}. As such, interfaces that promote recognition reduce the cognitive effort required from users. He suggests \textit{``Recognition rather than recall''} through his heuristics, meaning the user should not have to remember information from one part of the interface to another. %Other works began to approach this problem mainly from a technology standpoint by developing tools without considering the fundamental part of establishing and structuring good symbolism for sounds, resulting in bad choices for how sound was used. 
Approaching this problem from a purely technological standpoint by developing tools without considering the fundamental part of establishing and structuring good symbolism for sounds will likely result in misinformed choices about how sound is used.

In the visual realm, the most popular and comprehensive recommendation for assuring cognitively effective notations \cite{kuhar2021conceptualization} is \textit{The ``Physics''\ of Notations} (PoN) \cite{moody2009physics}. PoN suggests nine principles based on the perceptual properties of notations, %focusing on 
addressing and raising awareness about the importance of specific visual representation issues that software engineering researchers and notation designers have overlooked. 

Our current focus lies in establishing a solid groundwork to guide decision-making in creating an auditory notation. We leverage knowledge of music symbology and understanding the semiotics of the audible field, combined with the insights provided by PoN and other relevant research and tools. Our goal is to create a framework that can maximise the benefits of graphical notation for experienced software engineers and novices when combined with visual cues. Moreover, this approach presents the opportunity to model activities using only the aural sense, thereby promoting greater accessibility and inclusion for individuals with visual impairments in modelling activities. 

In this paper we introduce semiotics (Section \ref{sec:Semiotics}) and transpose Moody's work on graphical notational physics to sound (Section \ref{sec:revisting_physics}). Then, we review existing literature to establish design guidelines (Section \ref{sec:design_guidelines}). We then present our first proposal for a sound catalogue for UML Class Diagrams (Section \ref{cha:catalogue_sounds}) and report on an experimental study for its validation (Section \ref{sec:catalogue_study}) before concluding (Section \ref{sec:conclusion}).

\section{Semiotics} \label{sec:Semiotics}
To assemble an effective notation, we need to understand the influence different acoustic cues have on assigning meaning to sound. Semiotics is the study of any activity that involves the usage of signs and symbols and their signification. 
Generally, signs can be understood as a ``stand-in'' representation of a particular concept. Examples of signs include \textit{emojis} used in electronic communication, traffic lights and logos. Signs can also be drawings, paintings, photographs, words, sounds, or body language. The ones receiving the information need background knowledge to bridge the gap between the sign itself and the concept that it is meant to represent. 

In Semiotic Theory, the sign is made up of the relation between three components: the \textbf{(semiotic) object} - what the sign represents/encodes, and can be anything thinkable, for example, a fact or a law; the \textbf{signifier} - Also called the representamen, is the form that represents the denoted \textit{object}, giving its meaning. It can be, for example, a word or image and is also sometimes called representamen. Finally, the \textbf{signified} - also called interpretant, is the concept that a \textit{signifier} refers to. It is what is evoked in the mind - a mental concept \cite{chandler2007semiotics}. 
%Nota - o Peirce às vezes refere-se ao signifier como sign -> pode causar confusão porque aqui referimo-nos ao sign como o todo
Semioticians differentiate amongst various types of signs, emphasising that signs differ in how arbitrary/conventional they are \cite{kruselevels}. Regarding the signifier-object relations and how the signified denotes the object, 
we find philosopher Charles Peirce's second trichotomy \cite{everaert2019peirce} describing the different ways the sign refers to its object. As such, a sign can be an \textbf{icon} (firstness) - where the sign is perceived as resembling or imitating the object, being similar in having some of its qualities. %An icon may have as its signifier a qualisign, a sinsign or legisign. 
Examples of icons are sound effects in radio drama, a portrait or a scale model.

A sign can also be an \textbf{index} (secondness) - a mode in which the sign is affected by the object, the link between the signifier and signified can be observed or inferred, with both being directly connected in some way. This can be ascertained in examples like 'natural signs' - such as smoke, thunder, footprints or echoes; medical symptoms - pain, a rash, pulse rate, or 'signals' - like a knock on a door or a phone ringing.

Based on this brief theoretical explanation, along with the foundational work elaborated in Chandler's \textit{Semiotics: The Basics} \cite{chandler2007semiotics} and \textit{Conceptualization, measurement, and application of semantic transparency in visual notations} \cite{kuhar2021conceptualization}, %figure \ref{fig:semiotic_triangle} aims to aid the brief explanation of these concepts visually, while
Table \ref{tab:triadic_signs} details each of these terms, a concise semiotic meaning, a popular language meaning (how people colloquially perceive each of the terms), and a concrete example.

%\begin{figure}[htp]
%    \centering
%    \includegraphics[width=7cm]
%{Figures/SemioticTriangleFinal.jpg}
%    \caption{Semiotic Triangle}
%    \label{fig:semiotic_triangle}
%\end{figure}

\begin{table*}[hbt]
	\caption{meaning of triadic sign terms in semiotic theory and popular language}
	\label{tab:triadic_signs}
\centering
\begin{tabular}{|l|l|l|l|}
\hline
\textbf{Term} 	& 
 \textbf{Semiotic Meaning} & 
 \textbf{Popular language meaning} & 
 \textbf{Example}\\ \hline
	
Signifier & 
\begin{tabular}[c]{@{}l@{}} Is the form the sign takes, \\giving it meaning.\end{tabular} & \begin{tabular}[c]{@{}l@{}} The terms "representation" and \\"depiction" are used as synonyms.\end{tabular} & Drawing of a red heart \\ \hline

Signified &  
\begin{tabular}[c]{@{}l@{}} Is the concept that \\the signifier refers to.\end{tabular} &  \begin{tabular}[c]{@{}l@{}} The terms "concept" and "definition" \\ are used as synonyms. \end{tabular} & \begin{tabular}[c]{@{}l@{}} A symbol of love and affection\\ (Signifies the red heart)\end{tabular} \\ \hline

Sign & 
\begin{tabular}[c]{@{}l@{}}The unity of what is represented (the object), \\ how it is represented (signifier),\\ and how it is interpreted (signified). \end{tabular} & \begin{tabular}[c]{@{}l@{}} Object, quality, or event whose presence \\or occurrence indicates something else's \\probable presence or occurrence. \end{tabular} & Logos, Traffic lights\\ \hline

\end{tabular}
\end{table*}

%\section{Revisiting The ``Physics'' of Notations: Principles for an Auditory Notation} 
\section{Principles for an Auditory Notation} 
\label{sec:revisting_physics}

In this section, we transpose the principles proposed by Moody in a manner analogous to sound. We propose Table \ref{tab:principles_auditory_proposal} featuring the definition stated by Moody, this time applied to sound for each of these principles. The ``translation'' of each concept is not always direct, so the reasoning for each correspondence will be explained afterwards. 

%Tabela mais sintética comparação do physics of notations e som

%------------TABELA-------------------------------
\begin{table*}[hbt]
    \caption{The nine principles proposed by Moody with our equivalent auditory proposal.}
	\label{tab:principles_auditory_proposal}
\centering
\begin{tabular}{|l|l|l|}
\hline
\textbf{Principle} &
  \multicolumn{1}{l|}{\textbf{Definition}} &
  \multicolumn{1}{l|}{\textbf{Auditory Proposal}} \\ \hline
Semiotic Clarity &
  \begin{tabular}[c]{@{}l@{}}One-to-one correspondence between\\ symbols and their referent concepts.\end{tabular} &
  \begin{tabular}[c]{@{}l@{}}%Each sound should represent only one concept. \\ meaning that each of the proposed sounds \\ should only be played for only one concept.
  Each sound must have a single concept assigned to it, indicating that \\each proposed sound should be played for only one concept.
  \end{tabular} \\ \hline
Perceptual Discriminability &
  \begin{tabular}[c]{@{}l@{}}Different symbols should be \\ clearly distinguishable from one another.\end{tabular} &
  \begin{tabular}[c]{@{}l@{}}Different sounds should be clearly distinguishable from one another. \\For example, using different, unmistakable timbres for distinct sounds.\end{tabular} \\ \hline
Semantic Transparency &
  \begin{tabular}[c]{@{}l@{}}Visual representations whose\\ appearance suggests their meaning.\end{tabular} &
  \begin{tabular}[c]{@{}l@{}}Auditory representations should suggest their meaning. For example, \\the sound of a baby crying symbolises dependency.\end{tabular} \\ \hline
Complexity Management &
  \begin{tabular}[c]{@{}l@{}}Include explicit mechanisms to\\ deal with the complexity of a diagram.\end{tabular} &
  \begin{tabular}[c]{@{}l@{}}Noise of a container/box/door to signify ``composition'', supporting the \\concept of hierarchy. Usage of reverberation could be explored, a sound\\ with more echo should signify a further away diagram.\end{tabular} \\ \hline
Cognitive Integration &
  \begin{tabular}[c]{@{}l@{}}Include explicit mechanisms to support \\ integration of information from different\\ diagrams\end{tabular} &
  \begin{tabular}[c]{@{}l@{}}Using auditory motifs that refer to a specific diagram as a way\\ of contextualising it. 
  %Also, using TTS to orientate the user.
  In addition, it uses a text-to-speech component\\ to guide the user. 
  \end{tabular} \\ \hline
Visual Expressiveness &
  \begin{tabular}[c]{@{}l@{}}Use full range and capacities of visual\\ variables such as size, brightness, colour,\\ texture, shape and orientation.\end{tabular} &
  \begin{tabular}[c]{@{}l@{}}Auditory Expressiveness. Use the full range and capacities of auditory\\ variables such as timbre, pitch, loudness, duration, reverberation, etc.\end{tabular} \\ \hline
Dual Coding &
  Use text to complement graphics. &
  \begin{tabular}[c]{@{}l@{}}Certain sounds can be accompanied and complemented by a \\text-to-speech voice.\end{tabular} \\ \hline
Graphic Economy &
%  \begin{tabular}[c]{@{}l@{}}Number of different graphical symbols\\ should be cognitively manageable.\end{tabular} &
\begin{tabular}[c]{@{}l@{}}Number of different sound symbols\\ should be cognitively manageable.\end{tabular} &

  \begin{tabular}[c]{@{}l@{}}Finding the right balance between distinguishable \\sounds while being cognitively manageable.\end{tabular} \\ \hline
Cognitive Fit &
  \begin{tabular}[c]{@{}l@{}}Use different visual dialects for different\\ tasks and audiences.\end{tabular} &
  \begin{tabular}[c]{@{}l@{}}Creating diagrams using voice commands. Simple commands work for \\ beginners, while experienced users can use more complex ones.\\ Additional auditory cues may not be necessary for less experienced users.\end{tabular} \\ \hline
\end{tabular}
\end{table*}

%Semiotic Clarity
In the case of \textbf{Semiotic Clarity}, each sound should represent only one concept. This means that each proposed sound should be reserved for only one concept.

%Perceptual discriminability 
Different sounds should be distinguishable for the principle of \textbf{Perceptual Discriminability}. In the visual field, shapes play a special role in discriminating between symbols, representing the primary basis for identifying objects in the real world. Thinking about the characteristics that constitute a sound, the timbre (or ``tone colour'') can be considered its ``shape'', as it is a form of distinguishing different objects. Sounds should have various and unmistakable timbres to be recognisable. For example, a sound of an arrow being shot can be used for an \textit{association} in a UML Class diagram, while a baby crying can be used to describe a \textit{dependency}.

%Semantic Transparency
The successful exploration of this principle at the auditory level could potentially be more complex than its visual counterpart, namely when finding outstandingly intuitive sounds for ``abstract'' concepts. The auditory representations should suggest their meaning in \textbf{Semantic Transparency}. Neglecting this principle could result in semantically opaque sounds, with their meaning being purely arbitrary. Ideally, the sounds should be semantically immediate. Still, if they are semantically translucent (as mentioned above, the example of the baby crying symbolises \textit{dependence}), it would already be a great starting point for this research.

%Complexity Management
For the principle of \textbf{Complexity Management}, explicit mechanisms that deal with the complexity of a diagram should be included. For example, the noise of a container/box/door can signify ``composition'' and support the concept of a hierarchy. Reverberation could also be explored: a sound with more echo to signify a diagram that is further away. 

%Cognitive Integration - Podiamo-nos basear no Audible browser também, para "sumarizar" um diagrama
\textbf{Cognitive integration} must include explicit mechanisms to support integrating information from different diagrams. Using sound motifs to summarise the concepts of a diagram, Cognitive integration can be included. Sound motifs and text-to-speech (TTS) voice synthesis can also help contextualise directly related elements from other diagrams as foreign elements. Perceptual integration could be established using TTS, allowing for clear diagram labelling and easier way-finding on the users' behalf.

%Visual Expressiveness
The principle of \textbf{Visual Expressiveness} can be considered ``Auditory Expressiveness'', where auditory variables' full range and capacities are used. Krygier defines these variables as location, loudness, pitch, register, timbre, duration, rate of change, order and attack/decay \cite{krygier1994sound}. Besides these variables, reverberation can be helpful, as discussed in the principle of Complexity Management, or even panning the sound to the left or right channel.

%Dual Coding - acompanhar determinados sons com um Text-to-speech que serve para complementar o som
\textbf{Dual Coding} can be directly translated to an auditory approach. Text can be used to complement graphics in visual notation. A TTS voice can accompany certain sounds, reinforcing meaning as an additional cue to a given concept.

%Graphic Economy
The principle of \textbf{Graphic Economy} becomes ``Auditory Economy''. Just as graphical symbols should be cognitively manageable, so should the auditory ones, as these might be harder to assimilate in a first approach. One solution to this problem might be to increase auditory expressiveness by using multiple auditory variables (described in the principle of visual expressiveness) to differentiate between symbols. The right balance between distinguishable sounds while being cognitively manageable should be found. 

%Cognitive Fit
\textbf{Cognitive Fit} suggests the usage of different visual dialects when drawing and communicating certain aspects of a diagram. For an auditory approach, creating diagrams through voice can support simplified and more complex commands for novice and experienced users. While interpreting a diagram, certain auditory cues conveying additional information can also be omitted for less experienced users.

%------------------------------------------------------------------------%

\section{Design Guidelines} \label{sec:design_guidelines}

%"Pré Catálogo" - estabelecer umas "guidelines" para o que se pretende fazer 
%This section will detail and discuss already established guidelines and frameworks by published works. The goal is to facilitate the examination by documenting them all in one locale, corroborate some future decisions, and assist with streamlining certain aspects of this thesis' proposed work. 

%Towards a conceptual framework to integrate designerly and scientific sound design methods diz isto:
%In the phase of analysing a design problem, balancing acoustic and psychoacoustic aspects and psychological and socio-cultural dimensions is challenging. A particular challenge seems to be the multitude of social situations where sounds of everyday artefacts can be encountered. Many issues in the interaction with the client seem to lie in the arbitrariness of design argumentation. 

\subsection{Guidelines For Sound Design In The Context Of Software} 
%Blauert criteria for product sound design

Blauert \cite{blauert2005communication} argues that product sound design was initially considered unimportant and often treated as unwanted noise. Yet, he emphasises that product sound carries valuable information and can significantly improve the perceived quality of a product. Understanding the positive associations that certain sounds can create for a product is a key task in sound design. Despite the involvement of multiple sound designers in product planning, sound design still heavily relies on intuition rather than rational and systematic approaches. As a result, Blauert proposed various criteria for product-sound design:

%\begin{itemize}
%    \setlength{\itemsep}{1pt}
%    \setlength{\parskip}{0pt}
%    \setlength{\parsep}{0pt}
%    \item Motivation for product use - What are the reasons for product use? What does the listener gain when using the product? The product may compensate for deficiencies as a hearing aid does or make work much easier.
%   \item Function of the acoustic component in the product context - Is the acoustic signal a key quality element or just a secondary phenomenon in the background?
 %  \item Function of the acoustic component in relation to other product components - Are the sounds adequate considering other modalities, adverse conditions, etc.?
 %  \item Meaning associated with product use - e.g., applicability, suitability for use, saturation of demands, meeting of requirements, product performance, functionality, security.
  % \item Dominant quality items of the product's sounds - How is the product's sound judged? e.g. innovative, simple, complex, practical, distinctive, convincing or typical.
  % \item Provoked reactive behaviour of the user as to product quality - For example, the product is reliable, safe, high performance, creates emotional reactions such as disgust, contempt, resentment, dissatisfaction/satisfaction, empathy, etc., or motor activity like the ease of handling, ergonomy and flexibility.
  % \item User specification - Can a typical user be specified? If so, how can the group of users be typified, for example, according to age, education, experience of techniques and expectations of quality?
%\end{itemize}

\begin{itemize}
    \setlength{\itemsep}{1pt}
    \setlength{\parskip}{0pt}
    \setlength{\parsep}{0pt}
    \item Motivation for product use -  why is the product used and what does the user gain from it?
   \item Function of the acoustic component in the product context - is sound important to the product or just secondary?
   \item Function of the acoustic component in relation to other product components - are the sounds suitable for the product's use?
   \item Meaning associated with product use - how does the product meet user expectations?
   \item Dominant quality items of the product's sounds - how are the sounds judged?(e.g. innovative, simple, complex)
   \item Provoked reactive behaviour of the user as to product quality - how do users emotionally react to the product?
   \item User specification - can a typical user be specified, based on age, education, experience, or expectations?
\end{itemize}

Blauert \cite{blauert2005communication} believes that sound should enhance the product and meet users' needs. However, sound design has not received the same attention as visual design. Therefore, there is potential for more innovative sound design solutions.

%Sonification design guidelines to enhance program comprehension
 Hussein et al. \cite{hussein2009sonification} propose guidelines to address the challenges of combining visualisation and sonification in a single comprehension tool. They are as follows:

\begin{itemize}
    \setlength{\itemsep}{1pt}
    \setlength{\parskip}{0pt}
    \setlength{\parsep}{0pt}
    \item Add sonification to simplify visualisations - For example, a Call Hierarchy View in Eclipse. On a given method, Eclipse generates a tree visualisation of a call graph upon this request. However, this provides no information about the depth of the generated call graph. Adding a special visual cue that can show the depth of the tree is likely to clutter the tree visualisation. By contrast, adding a sonic cue representing the tree's depth via a different volume or pitch is relatively straightforward, minimising the required interaction time.
    \item Increase visual perception speed and accuracy by adding sonification - Auditory cues can enhance visual perception. E.g., multiple concurrent edits of the same source file complicate the subsequent merging of the changes. The authors argue that a sonic cue representing the number of concurrent edits can be rendered. This could effectively supplement the information already conveyed by a visualisation.
    \item Add sonification to present multiple information pieces simultaneously - %The use of sound 
    Sound use can improve comprehension and lower cognitive load when a person monitors various information sources updated concurrently. Musical scores convey lots of concurrent information together with their mutual relationships. For example, it could be used the attack-based multiple beats sonification for the metric of the number of lines of code. The "attack" refers to the initial part of a most noticeable and distinct sound, such as the sharp beginning of a drumbeat serving as a trigger to begin playing the next sound in the sequence. Multiple beats sonification means multiple sounds or beats played in a pattern or sequence where each sound represents a specific data point or information. The sequence of sounds creates a sonic representation of the data.

%Multiple beats sonification means multiple sounds or beats played in a pattern or sequence. Each sound represents a specific data point or piece of information, and the sequence of sounds creates a sonic representation of the data.
    \item Use sonification to summarise information - A simple sonification could effectively complement a visualisation by summarising large volumes of information. The authors give the example of using a sonic cue to express a relative length or the cyclomatic complexity metric of a selected source file.
    \item Interchange visualisation and sonification to improve effectiveness - Although audio may not necessarily convey more information, it covers more ground spatially and, as such, could provide more distinguishable entry points for monitoring. The authors state that a study has observed that sound superseded visuals in terms of the conveyed detail, while at other times, the situation was the opposite. This insight suggests that aural cues could be more effective than their visual counterparts when utilised in the same scenario under certain circumstances.
    \item Alternate visualisation and sonification to improve accessibility - The importance of accommodating users with disabilities has been widely recognised. Auditory representations of the information could supplement the otherwise inaccessible cues for impaired users.
\end{itemize}

% There's More to Sound Than Meets the Ear: Sound in Interactive Environments - Sound Design
Kenwright \cite{kenwright2020sound} states that audio is often an afterthought in many interactive environment projects, stressing that, like all aspects, the sound should be designed. The author gives examples of how sound design fails in the context of serious games. The main ones listed below are adapted into a broader scope that can encapsulate in our work.

%{\small
\begin{itemize}
    \setlength{\itemsep}{1pt}
    \setlength{\parskip}{0pt}
    \setlength{\parsep}{0pt}
    \item Positional audio appears nonexistent - Difficulty to tell where voices are coming from
    \item Volume mixing is inconsistent and random - Actions have a soft sound, and voice audio is too low
    \item Actions sound weak in comparison to everything else - Some sounds are barely audible, while subsequent sounds may be too loud
    \item Actions with no corresponding audio - Something happens, but there is no audio indication
    \item Sound which contradicts one another in tone and style
\end{itemize}
%}

In the context of this work, the examples listed above could be extrapolated to having positional audio representing the different positions where elements in a diagram are located. Since these elements can be located more to the left, right, higher or lower, the sound representing these elements should translate this positioning into a 2D plane. In addition,
if there is a TTS reading of the various elements that make up the diagram, the voice sound must have a balanced volume and sounds that represent the sound notation. 

%SUBSECTION- SCIENTIFIC METHOD FOR SOUND DESIGN
%Towards a conceptual framework to integrate designerly and scientific sound design methods
\subsection{Sound Design Scientific Method: A Conceptual Framework} \label{sec:scientific_SoundDesign_framework}
Hug and Misdariis \cite{hug2011towards} developed this framework of concepts and heuristics to help inform design decisions. The framework is divided into three main components (Typology Of Interactive Commodities; Situational Heuristics, and Narrative Metatopics), subdivided into multiple categories. 

%Typology Of Interactive Commodities
\textbf{Typology Of Interactive Commodities} was developed along with degrees of abstraction of sound and object. As Hug and Misdariis state, sound is closely related to physical and material processes. It plays a core role in communicating an object's ``hidden'' qualities, such as its stability or solidity. This component is intended to help orient the sound design strategy used and is divided into the categories:

\begin{itemize}
    \setlength{\itemsep}{1pt}
    \setlength{\parskip}{0pt}
    \setlength{\parsep}{0pt}
    \item \textbf{Authentic Commodities} - Simple, self-contained, fitting with existing sonic identity
    \item \textbf{Extended Commodities} - Sound not necessarily related to object's sonic identity, communicates extension quality
    \item \textbf{Placeholders} - Proxies of a virtual object. Sound defines the virtual object. For example, the Wiimote or Tangible User Interfaces
    \item \textbf{Omnivalent Commodities} - Sound defines the artefact. Defined through software rather than physical configuration
\end{itemize}

%Situational Heuristics
\textbf{Situational Heuristics} concern the situational categories that define the relationship between interactive commodities and their use context. These categories are:

\begin{itemize}
    \setlength{\itemsep}{1pt}
    \setlength{\parskip}{0pt}
    \setlength{\parsep}{0pt}
    \item \textbf{Social Situation} - private; public
    \item \textbf{Level Of Intimacy} - objectified (meaning: totally detached from human body); pocketable; wearable; implant
    \item \textbf{Relationship To User and Task} - assistant; tool
    \item \textbf{Type Of Use} - casual; professional
\end{itemize}

\textbf{Narrative Metatopics} are abstracted themes and attributes associated with narratively significant artefacts and interactions in fictional media, like films or games \cite{hug2011towards}. These provide a means of navigating a complex semantic space and can be associated with a collection of specific sound design strategies, which, according to the authors, serve to build grounded sonic interaction design hypotheses as a starting point for the design. %They are also meant to link qualities of interactive %processes with qualities of sonic processes. These are:

\section{Sounds Catalogue for UML Class Diagrams} \label{cha:catalogue_sounds}

We propose a sounds catalogue for Class diagrams in UML, based on auditory principles and semiotics. The aim is to improve the perception of sounds for users, compared to previous efforts with arbitrary choices.

To define our Catalogue, we start from the basis of semiotics, which tells us that a sign is composed of the relationship between the signifier, object and signified. The chosen sound cues are our \textit{signifiers} since they are the representation that our signs will have. Looking at the table that maps the semiotics of the audible field, the sounds in our Catalogue all fall within the Aesthetical Regime, being considered ``Soundtracks'', as these tend to offer a wide range of articulated sound effects. Our sound cues can be regarded as auxiliary legisigns (a type of sign created by a law or convention) replicated to collaborate with signifying processes mainly conveyed through the visual form, such as the case with the visual nature of UML diagrams. Our \textit{objects} are the UML elements in question, which are the concept each of the chosen sounds will represent. Finally, the \textit{signified} defines a given UML concept.

\begin{figure}[htp]
    \centering
   \includegraphics[width=7cm]{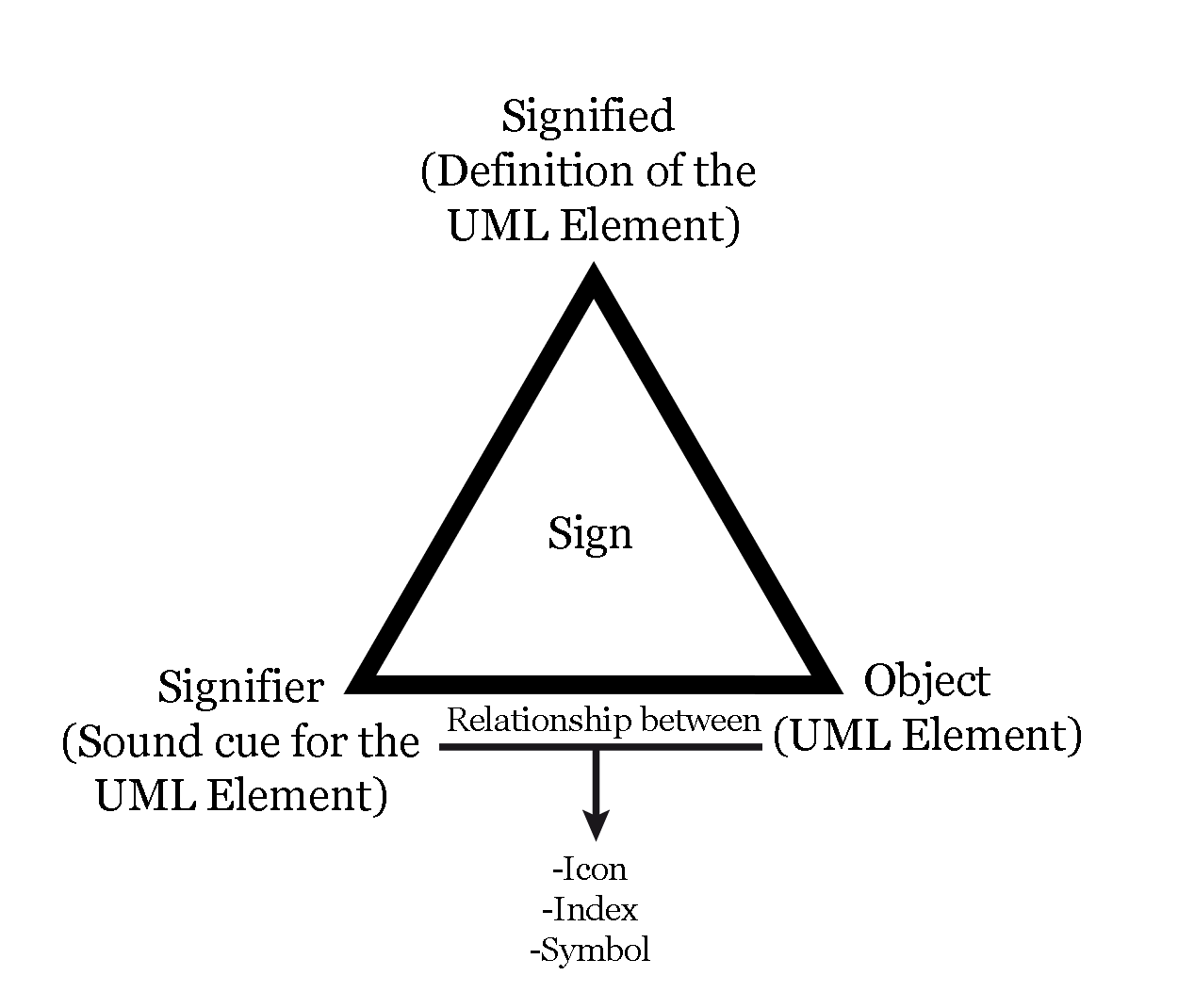}
   \caption{Semiotic Triangle For The Catalogue Of Sounds} 
    \label{fig:semiotic_triangle_catalogue}
\end{figure}

A sound cue and its relation to the UML element it represents can be considered an Index, an Icon or a Symbol. Figure \ref{fig:semiotic_triangle_catalogue} %is based on figure \ref{fig:semiotic_triangle} and
%concisely illustrates what has been described, denoting 
denotes
the relationships between the different elements that constitute a sign in this specific context. %It should also be mentioned that the 
%The form in which the signified denotes the object (the third trichotomy) will not be explored in this paper, as UML concepts are well-established in the MODELS community. %the definition of a UML concept is already well established within UML documentation.

%FIGURA Semiotic Triangle

%--------------------------------------------------------%
Our catalogue follows the auditory principles proposed in this document. More precisely, it follows the principles of semiotic clarity and semantic transparency, along with perceptual discriminability, auditory expressiveness and economy. As this Catalogue only focuses on defining sounds for the basic elements that form a UML class diagram, the principles of cognitive integration and dual coding won't be applied here, as these concepts imply more complex relationships between the different elements in a UML diagram, along with the need for the usage of a TTS voice.

All sounds apply the principle of auditory expressiveness since they use the full range and capacities of auditory variables. For example, the pitch represents this relationship between classes in the inheritance sound. The principle of auditory expressiveness is also completely applied as we use the various characteristics that make up a sound. Furthermore, all the sounds created for the experimental study and subsequent use in the tool developed took into account the guidelines previously detailed. These sounds all have a maximum of 3 seconds, considered brief auditory icons (customarily named ``earcons'') to be more easily recognisable by users. Furthermore, %it should be noted that 
the chosen sounds also try to accommodate the universal human experience to be as understandable as possible, as there may be certain sounds that people from distinct cultures and backgrounds will interpret in different forms.

Table \ref{tab:catalogue_good} further expands on these concepts by featuring each diagram element, its definition, the defined sound and the reasoning behind the choice.

\begin{table*}[hbt]
    \caption{Catalogue Of Sounds for UML Class Diagram Elements} 
    \label{tab:catalogue_good}
\centering
\begin{tabular}
{|m{0.08\linewidth}|m{0.13\linewidth}|m{0.72\linewidth}|}
%{|m{0.056\linewidth}|m{0.281\linewidth}|m{0.094\linewidth}|m{0.51\linewidth}|}
\hline

%\begin{table*}{|m{0.056\linewidth}|m{0.281\linewidth}|m{0.094\linewidth}|m{0.51\linewidth}|}%|>%{\hspace{0pt}}m{0.056\linewidth}|>{\hspace{0pt}}m{0.281\linewidth}|>{\hspace{0pt}}m{0.094\linewidth}|>{\hspace{0pt}}m{0.51\linewidth}|} 
\textbf{Diagram Element} & \textbf{Sound} & \textbf{Explanation} \\ 
\hline
Class & Book opening + positive/correct sound & Opening a book emits a high-pitched, ``bright'' sound that signals the beginning of accessing more information about attributes and operations. The rising pitch of the sound captures the user's attention and implies that something important is happening, often interpreted as a positive signal.\\
 %A rectangle represents them with rows of the class name, attributes, and operations. & Book opening + positive/correct sound & Opening a book signifies the opening of something that contains more information regarding attributes and operations. The correct sound is high-pitched, so it sounds "bright". The note becoming higher pitched indicates to the user that something is happening, which is an important sound, leaving the user more attentive to something that could have more information. The fact that it is a sound that goes higher in pitch is also normally associated with a "positive" sound. \\ 
\hline
Attribute & Wooden bricks falling + computer notification sound (beeps) & %The sound of wooden bricks falling symbolises the building blocks that will be used to build something in this case, the attributes of a class can be seen as the building blocks of a class.The sound of the notification, through its beeps and sequence, is a sound that points towards something "digital", which represents information and data.The combination of these two sounds points the users to the notion of the attribute. 
The sound of falling wooden bricks symbolises the building blocks used to construct something, such as the attributes of a class. Similarly, the digital notification sound conveys information and data. Together, these sounds evoke the idea of an attribute to users.
\\ 
\hline
Operation / Method & Keyboard Typing & %An operation is linked to the act of programming, so the typing sound symbolises the act of programming the specification of a method \\ 
The typing sound symbolises the act of programming a method's specification, as an operation is inherently connected to programming.\\
\hline
Association & Arrow shot & %Symbolises a relationship between two or more classes. We can associate the sound of an arrow being shot to a cupid, which shoots arrows. Generally, the cupid symbolises an association/relationship between people. Furthermore, an arrow is also used to represent this concept in UML diagrams visually, so it makes sense to have this equivalence for sound \\ 
An arrow being shot symbolises a relationship between two or more classes, similar to how Cupid shooting arrows represent an association between people. The arrow is also used to visually represent this concept in UML diagrams, making it fitting to have an equivalent sound.\\
\hline
Inheritance & Book opening (lower pitched) + Coins falling + Book opening (higher pitched) & %It's symbolised by a straight connected line with a closed arrowhead pointing towards the superclass. & The sound of the book opening is a motif that points to the sound used for classes. The higher and lower pitched sound is symbolic of the human voice's higher and lower pitched voice, which in turn is related to the sound of a parent and child, a concept related to Inheritance. The pitch usage in this sound also respects the principle of auditory expressiveness. Finally, the sound of falling coins tries to make more evident that this is the concept of Inheritance, which is usually associated with wealth and, therefore, money. These three sounds combined represent the "passing down" of Inheritance (represented by the sound of coins), from the lower-pitched sound for a class to the higher-pitched sound for a class \\ 
The sound of a book opening is a motif that signifies the sound for classes, using higher and lower pitches that reflect the range of the human voice, evoking the parent-child relationship of Inheritance. The pitch choice aligns with the principle of auditory expressiveness. The falling coins sound strengthens the association of Inheritance with wealth, and the combination of these three sounds represent the passing down of Inheritance from a lower-pitched sound for a class to a higher-pitched sound for a class.\\
\hline
Realisation / Implementation & Construction noise in the background with a hammering sound standing out & %The ambience of construction noise in the background already takes our minds to a place where something is being built. Having the hammer sound stand out makes what is depicted slightly clearer. As mentioned, these sounds represent something being built/implemented, which is what is intended to be represented with this concept. \\ 
The background noise of construction sets the stage for building/implementation, while the sound of a hammer emphasises the idea of construction. Together, these sounds represent building and implementing, conveying the intended concept.\\

\hline
Dependency & Baby crying & %The sound of a baby crying symbolises the infant's dependency on its parents. This sound follows the principle of semantic transparency since the auditory representation suggests the meaning intended for this concept. \\
The sound of a baby crying represents the dependency of infants on their parents, aligning with the principle of semantic transparency by conveying the intended meaning through auditory representation.\\
\hline
Aggregation & Sports crowd & %Crowd noise symbolises an aggregation of people, which can be associated to the concept of aggregation, in which objects are assembled to create a more complex object. This particular sound points the listener to a crowd at a sports game, where this concept can also be associated with a team composed of players. If a team is disbanded, the players live on. This, in turn, meets the concept represented in aggregation, which states that destroying the whole does not eliminate the parts. \\ 
Crowd noise symbolises an assembly of people, associated with the concept of aggregation where objects combine to form a more complex object. The sound evokes the image of a sports game crowd, related to a team of players. Disbanding a team doesn't eliminate the players, meeting the principle of aggregation where destroying the whole doesn't remove the parts.\\
\hline
Composition & Sports crowd + fire burning & %Just like aggregation, the crowd noise symbolises a gathering of people, which turns the users to the concept of aggregation. Because composition is a strong form of aggregation, it makes sense to have this sound as a motif that reminds the users of this concept. However, the principle of semiotic clarity tells us that each sound should only represent one concept, so the composition has a fire-burning sound to differentiate these two concepts further. This is a destructive sound, which indicates that in this concept, the parts are destroyed along with the whole. \\ 

The gathering of people represented by crowd noise draws attention to the concept of aggregation, which is related to composition. To distinguish between these two concepts, the fire-burning sound is used to represent composition, following the principle of semiotic clarity by assigning only one concept to each sound. The destructive nature of the fire-burning sound conveys that, in composition, the parts are destroyed along with the whole.\\
\hline
Association Class & Arrow sound + Book opening + pages in a book sound & This sound associates the concept of the association relationship and the concept of class. For this matter, it seems appropriate to reuse the already-defined sounds. The arrow sound symbolises the association, joined with the defined sound for the concept of class (book opening) and the new sound of the pages inside the book. This new sound indicates to the user that this is a new sound. In this way, every sound is distinct for each class diagram element, obeying the principle of perceptual discriminability, but the user is reminded of the concepts already established before (semiotic clarity is respected, seeing that these sounds only represent one concept) \\ 
\hline
Package & Envelope being opened + zip opening & %Follows the same idea as the sound is chosen for the class concept. 
The envelope being opened ties into the sound used for the concept of a class (book opening) through the usage of paper. An envelope is not only a piece of paper but also usually contains written information inside it. Furthermore, an envelope can symbolise mail (visual icons for this concept usually represent it through a closed envelope) and can be associated with delivering ``packages''. The sound of the zip of a bag being opened represents something that contains more information while also being reminiscent of a mailman's bag. \\
%This sound combines the concepts of association and class, utilising previously defined sounds for clarity. The arrow sound represents association, while the book opening sound symbolises class. To differentiate this new sound, the pages inside the book sound are used, emphasising distinctiveness for each class diagram element as per the principle of perceptual discriminability. However, semiotic clarity is maintained by assigning only one concept to each sound while reminding the user of established concepts.\\

\hline
\end{tabular}
\end{table*}

%TODO? - Completar um pouco melhor esta parte, mencionar de forma um pouco mais completa como é que são cumpridos os principios?
%Semiotic Clarity - each sound should only represent one concept
%Semantic transparency - auditory representations should suggest their meaning -> este ponto vai de encontro à semiotica e ao simbolismo também

%Perceptual discriminability -different sounds should be cleary distinguishable from one another
%Auditory expressiveness - usar timbre, pitch, loudeness, duration, reverberation
%Auditory economy  - número de diferentes sons ser gerível -> este ponto com este catálogo será sempre cumprido, uma vez que número de elementos dos diagramas é limitado

%-------------%
%Na ferramenta
%Cognitive integration - auditory motifs (só seria possivel ser incorporado na ferramenta, mas não será bem testado)
%Dual coding - envolve tts (só será possível ser incorporado na ferramenta)

%Estes dois pontos infelizmente não há facilidade em fazer algo desse genero. Teria de se fazer com outras linguagens que não UML e isso já saíria do escopo do documento
%Complexity management
%cognitive fit 

%--------------------Experimental Study-----------------------
\section{Experimental Study} \label{sec:catalogue_study}
%This section presents the experimental procedure regarding the Catalogue of sounds for UML Class Diagrams.

\subsection{Planning}
%Before the experiment was put into practice, it was necessary to plan all the details leading up to it. The experimental study has been structured in the phases presented in the following sub-chapters.

\subsubsection{Goals}
We describe our goals following the GQM template \cite{Basili1994GQM}. The first goal (\textbf{G1}) is to \textit{analyse} the difference between using a sound catalogue built according to the proposed auditory principles and the semiotics of the audible field, and a catalogue that disregards said principles and semiotics, with an arbitrary choice of sounds, \textit{for the purpose of} their evaluation \textit{with respect to} user preference on those sounds, \textit{from the viewpoint of} researchers, \textit{in the context of} an experiment conducted with Computer Science graduate students and professionals. Our second goal (\textbf{G2}) is to \textit{analyse} the adoption of the proposed auditory principles, \textit{for the purpose of} their evaluation, \textit{with respect to} their relevance to users, \textit{from the viewpoint of} researchers, \textit{in the context of} an experiment conducted with Computer Science graduate students and professionals.

%A study was conducted to qualitatively determine if the construction of a catalogue of sounds following a more systematic approach based on the proposed auditory principles and the semiotics of the audible field is more adequate and effective when compared to a catalogue that disregards said principles and semiotics, where the choice for sounds is purely arbitrary. Additionally, we want to determine how relevant the previously proposed auditory principles are to the users.

\subsubsection{Participants}
We looked for participants who were either professional software engineers or graduate and undergraduate Computer Science students. Participants needed basic knowledge of English, a device capable of sound output, and a reliable internet connection to model with UML class diagrams and evaluate sounds through an online form.
We recruited 31 participants through convenience sampling. We leveraged personal contacts and invited participants through direct contact or e-mail. 26 participants were male, and the remaining 5 were female. 2 of the participants reported being hard of hearing. Concerning their self-reported level of expertise, 4 participants rated themselves as \textit{excellent}, 10 as \textit{very good}, 14 as \textit{good}, 3 as \textit{average}, and 0 as \textit{poor}. 11 of our participants use UML regularly, 9 have not used UML for 1 year, and the remaining 11 have not used UML for at least 2 years.

%Participants needed basic knowledge of English, a device capable of sound output, and a reliable internet connection to model with UML class diagrams and evaluate sounds through an online form. Convenience sampling was used to select participants among computer engineers and master and undergraduate students in Computer Science.

%Questions were asked to understand better participants' profiles and level of knowledge as well as gender, hearing problems, and expertise in UML.

%\begin{itemize}
%\item Gender
%\item Are you hard of hearing?
%\item How do you consider your level of understanding of UML?
%\item When was your last contact with UML?
%\item In what context do you use UML?
%\end{itemize}

%In total, this experimental study had 31 participants.

\subsubsection{Experimental materials}
We created two sound catalogues, the proposed catalogue, following auditory principles and semiotics (see Table \ref{tab:catalogue_good}) and a baseline catalogue. 

For the baseline catalogue, we selected sounds that violated auditory principles and semiotics. We made arbitrary choices for all other aspects. For each concept in UML class diagrams, we assigned a sound and an explanation of how it failed to satisfy the proposed auditory principles. Table \ref{tab:catalogue_bad_practices} presents the baseline catalogue, listing a sound and its corresponding unsatisfied auditory principles for each diagram element.
{\small
%TABELA CATALOGO DE UNSATISFIED AUDITORY PRINCIPLES
\begin{table*}[!htb]
\centering
\caption{The Catalogue Of Bad Practices with the Unsatisfied Auditory Principles}
\label{tab:catalogue_bad_practices}
\begin{tabular}{|p{0.13\textwidth}|p{0.25\textwidth}|p{0.55\textwidth}|} 
\hline
\textbf{Diagram Element} & \textbf{Sound} & \textbf{Unsatisfied Auditory Principles} \\ 
\hline
Class & Sound of a car engine & Semiotic Clarity (along with Attribute); Semantic Transparency \\ 
\hline
Attribute & Sound of a car engine & Semiotic Clarity (along with Class); Semantic Transparency \\ 
\hline
Operation / Method & Sound of running water & Semiotic Clarity (along with association); Perceptual Discriminability (along with Association); Semantic Transparency \\ 
\hline
Association & (A different) sound of running water & Semiotic Clarity (along with Operation/Method); Perceptual Discriminability (along with Operation/Method); Semantic Transparency \\ 
\hline
Inheritance & Farm animals + piano notes + window being cleaned + tyres breaking + plastic bottle being crushed + the sound of a car engine & Semiotic Clarity; Perceptual Discriminability; Semantic Transparency; Auditory Economy \\ 
\hline
Realization / Implementation & Sound of the wind & Semiotic Clarity (Along with Dependency); Semantic Transparency \\ 
\hline
Dependency & (A different) sound for the wind & Semiotic Clarity (Along with Realization/Implementation); Semantic Transparency \\ 
\hline
Aggregation & Elephant Sound & Semantic Transparency \\ 
\hline
Composition & Cartoon running Sound & Semantic Transparency \\ 
\hline
Association Class & Doorbell Sound & Semantic Transparency \\ 
\hline
Package & Explosion Sound & Semantic Transparency \\
\hline
\end{tabular}
\end{table*}
}

The choice of a car engine sound for the Class and Attribute elements was made due to their violation of the Semiotic Clarity and Semantic Transparency auditory principles. Using the same sound for both violates Semiotic Clarity, and the sound fails to convey these concepts' meaning, violating Semantic Transparency. This results in a Symbol where the signifier does not resemble the signified.

The sound of running water was chosen for the elements' Operation/method and Association, despite the two different sounds representing these elements. This violates the principle of Perceptual Discriminability since the two sounds are very similar and indistinguishable. In contrast to the Class and Attribute elements, where the same sound was used for both concepts, semiotic clarity is not satisfied here because the same symbol is used for two different concepts. Additionally, running water does not suggest these concepts' meaning, violating the Semantic Transparency principle.

The sound chosen for the Inheritance element is a mixture of several sounds, including farm animals, piano notes, windows being cleaned, tyres breaking, plastic bottles being crushed, and the sound of a car engine. However, this violates several auditory principles. The principle of Auditory Economy is not satisfied, as the number of auditory symbols used becomes too much to handle. The sound also violates the principle of Semantic Transparency, as it does not suggest the concept's meaning. Furthermore, Perceptual Discriminability is not satisfied as a sound of a different car engine makes it indistinguishable from the Class and Attribute sound. Semiotic clarity is also not satisfied, as despite the car sound being different from the others, it is still the same symbol for a different concept. Additionally, this sound is too long, which goes against the guidelines outlined in Chapter \ref{sec:design_guidelines}: \textit{``use sonification to summarise information''}.

Two different sounds of wind were chosen for the elements of \textit{Realisation/implementation} and \textit{dependency}. This means that these do not satisfy the principle of Semantic Transparency since the wind does not suggest the meaning of any of these concepts. The principles of Semiotic Clarity and Perceptual Discriminability were also disobeyed for the same reasons as seen before with the elements \textit{Operation} and \textit{Association}, meaning that the same symbol is used for two different concepts, and two distinct sounds are too similar and not distinguishable, respectively.

Finally, for the elements \textit{aggregation}, \textit{composition}, \textit{association class} and \textit{package}, sounds that violate the principle of Semantic Transparency were chosen. In addition to the principles that were not followed in this Catalogue, the sounds subsequently created to represent what is detailed here do not follow the guidelines explored in section \ref{sec:design_guidelines}, with examples such as: \textit{``volume mixing inconsistent and random''} and \textit{``sound which contradicts one another in tone and style''}. The junction of cartoon sounds with real-life sounds exemplifies the latter. For example, the sound of a car (used to represent a \textit{class} and \textit{attribute}) and the cartoon sound used for the concept of \textit{composition}, or the variety of sounds used for \textit{inheritance}, makes this catalogue vary in tone and style dramatically.

We selected basic sounds from \textit{zapsplat.com} and edited them using \textit{REAPER}~\footnote{\url{https://www.reaper.fm/}} to create the sounds for both catalogues. %See Figure \ref{fig:reaper_zapsplat} for an example.

% REPOR QUANDO ACEITE: 
Both sound catalogues and the evaluation results are available on this paper's companion site at Zenodo\footnote{\url{https://doi.org/10.5281/zenodo.7766679}}.

%We used Google Forms to build a questionnaire for our evaluation session. Each question focused on one element of UML Class diagrams and included a visual representation of the element, a brief definition and the two corresponding sounds, one from each catalogue. The sounds were available via links embedded in the form. To mitigate the risk of bias the order of presenting the sounds was randomised: for some questions, the first sound was the baseline, and the second one was the sound in our proposed catalogue, while for others it was the other way around. 

We created a questionnaire for our evaluation session using Google Forms. Each question featured a visual representation and a brief definition of one UML Class diagram element, along with two sounds - one from each catalogue - which were accessible through embedded links. To avoid bias, the presentation order of the sounds was randomized, with some questions presenting the baseline sound first and the proposed sound second, while others presented them in reverse order.

%Sounds created with \textit{REAPER} were added to the form by uploading them to Google Drive and embedding links in the form. The links were made accessible by selecting the "any person with the link" option under "General Access" in the Share button.

\subsubsection{Tasks}
%Participants had to answer a form where they were presented with the elements of a UML class diagram and their visual representation. For each of these concepts, users were then asked to listen to the sound previously defined in the Catalogue that follows good practices and a sound from the Catalogue that doesn't satisfy the auditory principles. Subsequently, the respondents were asked to select which sound they thought best suited the concept being presented to them, being able to explain their choice by text if they so desired. If the participant didn't think the two sounds matched the concept, they could suggest their own and explain it.

%After this task, participants had to answer a few additional questions to evaluate the relevance of the previously proposed auditory principles.

\begin{figure}[htp]
    \centering
   \includegraphics[width=9cm]{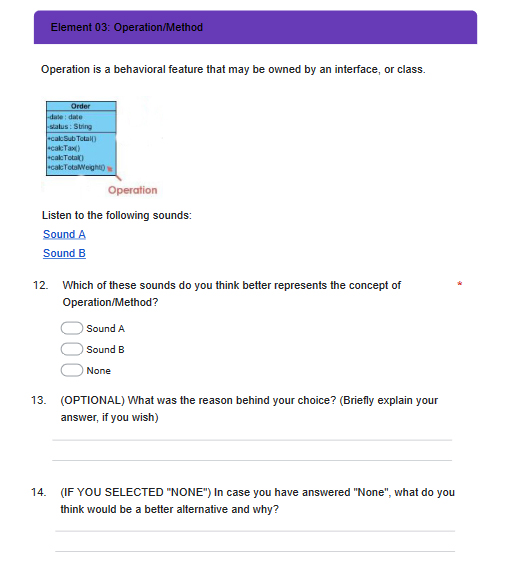}
   \caption{Structure of a typical set of questions in the form.}
    \label{fig:question_form}
\end{figure}

Participants filled in a form where they could observe UML class diagram elements, their visual representation, and a short definition. Then, they listened to two sounds, one from the proposed catalogue, and the other from the baseline catalogue. They selected the best sound for each concept and could provide text explanations. If they did not like either sound, they could suggest their own. Figure \ref{fig:question_form} illustrates the form for the \textit{operation} model element. 

After selecting their preferred sounds for each UML element, participants were presented with the definitions of the proposed principles (as seen in Table \ref{tab:principles_auditory_proposal}). They were then asked to rate how relevant they thought each of those principles was for the definition of adequate sounds to represent the different UML model elements. They were also asked a question about the overall relevance of these principles. Finally, we also asked them for demographic information.

\subsubsection{Hypotheses and variables}
For each of our high-level goals (G1, G2), we define a null hypothesis and the alternative hypothesis. G1 concerns the preference expressed by participants towards choosing a sound selected according to the proposed auditory principles and the semiotics of the audible field \textit{vs.} choosing a baseline sound chosen in disregard for said principles and semiotics.
{\small
\begin{framed}
\noindent $H_{0\mathit{Catalogue}}$: There is no difference in terms of \textbf{preference} in using sounds from the catalogue built following the proposed auditory principles and semiotics of the audible field and using sounds from a baseline catalogue built disregarding said principles and semiotics for UML model elements.

\noindent $H_{1\mathit{Catalogue}}$: There is a significant difference in terms of \textbf{preference} in using sounds from the catalogue built following the proposed auditory principles and semiotics of the audible field and using sounds from a baseline catalogue built disregarding said principles and semiotics for UML model elements.
\end{framed}
}
G2 concerns the perception of the relevance of auditory design principles and semiotics of the audible field in designing a catalogue of sounds for UML class model elements, as proposed in this paper.
{\small
\begin{framed}
\noindent $H_{0\mathit{Principles}}$: There is no difference in terms of \textbf{perceived relevance} in using the proposed auditory principles to design a catalogue of sounds for UML class model elements as proposed in this paper.

\noindent $H_{1\mathit{principles}}$: There is a significant difference in terms of \textbf{perceived relevance} in using the proposed auditory principles to design a catalogue of sounds for UML class model elements, as proposed in this paper.
\end{framed}
}
The \textbf{independent variables} in both cases are the \textit{catalogue} and the \textit{model element}. This is a \textbf{nominal} variable which can assume one of two values, \textit{proposed} and \textit{baseline}. 
The \textbf{dependent variables} are:
\begin{itemize}
    \item \textbf{G1}: the expressed \textit{preference}, measured in a nominal scale value for each catalogue and model element, with one of three possible values - \textit{Sound with Rationale}, \textit{None}, and \textit{Sound without Rationale}. A higher concentration of choices favouring \textit{Sound with Rationale} would support the claim on the preference for sounds leveraging the proposed auditory principles and semiotics of the audible field.
    \item \textbf{G2}: the \textit{perceived relevance} of each design principle, measured as a 5-point Likert scale ranging from \textit{Strongly disagree [with the relevance of the design principle]} to \textit{Strongly agree [with the relevance of the design principle]}. Higher values in this Likert scale would support the claim for a high perceived relevance of the design principles proposed in this paper.
\end{itemize}

\subsubsection{Experimental design}
%We followed a within-subjects design, where participants were subjected to both treatments. In other words, for each model element, participants listened to both audios. We opted for this so each participant would always be able to compare sounds from both catalogues. This was also advisable given the relatively scarce availability of candidate subjects.
We employed a within-subjects design, where each participant was exposed to both treatments. This allowed for a comparison of sounds from both catalogues for each model element, which was important due to the scarce availability of candidate subjects.

\subsection{Execution}

\subsubsection{Preparation}
We prepared all experimental material beforehand. This included developing both catalogues (the proposed and the baseline) and setting up the questionnaire in Google Forms. Then, we ran a pilot with a junior member of our research team, to assess and refine our data collection instrument and gauge the estimated time to complete the full task. As participation in the experiment was built around that questionnaire, we made sure it was available online to anyone with the link. To prevent multiple submissions, respondents were required to log in with their email.

\subsubsection{Procedure}
The evaluation sessions were conducted online, with a single participant each time. Participants were free to use any device of their preference as long as it was capable of producing sound output and had a visual display to access the form. We used a personal computer with headphones and a good internet connection.
%Before starting the experiment, a brief explanation was presented along with a test sound to ensure the participant's setup was capable of emitting audio. We then asked a few demographic questions, to better characterise our participants. This was followed by the evaluation itself, where participants would go through the Google Form and evaluate the sound proposals for each of the UML class modelling elements covered in this study. Participants listened to two sound options via hyperlinks to Google Drive sound files and selected which better represented the UML concept, with the option to provide an explanation of why they preferred that sound or to recommend a better alternative.
A brief explanation and a test sound were provided before the experiment. Demographic questions were asked before participants evaluated the sound proposals for each UML class modelling element covered in the study. Participants listened to two sound options via hyperlinks to Google Drive sound files and selected which better represented the UML concept, with the option to provide an explanation or recommend a better alternative.

Participants rated the importance of each proposed principle on a scale of 1 to 5 and their overall relevance, as shown in Table \ref{tab:principles_auditory_proposal}, after evaluating the sounds. This provided insight into how participants perceived the importance of the principles in their responses. On average, each session lasted about 30 minutes.

%After answering these questions, we asked participants to rate the importance of each proposed principle on a scale of 1 to 5 and to share their opinion on the overall relevance of the principles, as shown in table \ref{tab:principles_auditory_proposal}. This helped to assess how participants perceived the importance of the principles in relation to their responses. On average, each session took around 30 minutes.

\subsubsection{Deviations from the plan}
The experimental procedure was conducted according to the plan. That said, due to a limitation in the questionnaire, visible in Figure~\ref{fig:question_form}, question 14, participants could answer some questions even when they did not need to (e.g. some participants would answer question 14, even if they chose Sound A or Sound B). Those unsolicited answers were discarded from our analysis.

\subsection{Analysis}
\subsubsection{Data set preparation}
%The data collected during the experiment using the Google Form was saved as a spreadsheet. We imported the data concerning the choices of sounds and the perceived relevance of the different principles for auditory notations into a statistics tool (SPSS) which we used for further analysis.
The experiment data, collected through a Google Form, was saved as a spreadsheet and imported into the statistics tool (SPSS) for analysis. This included the data on sound choices and the perceived relevance of principles for auditory notations.

\subsubsection{Analysis procedure}
We started by performing a frequency analysis for the preference of sounds for each model element, and the perceived relevance of each principle. We also collected descriptive statistics for the latter, including the number of cases, mean, standard deviation, minimum and maximum. We then conducted, for both cases, a Chi-Square Goodness of Fit Test, with the assumption that the expected differences would be similar for all groups. This assumption means, for G1, that there would be a similar likelihood of preferring the sounds from the proposed catalogue, the baseline, or none of them. For G2, this assumption would mean that the principles are essentially indifferent when choosing adequate sounds. We used $p < 0.05$ for the level of significance. As we had several related tests within G1 and within G2, it can be argued that this implies a form of multiple comparisons. The Bonferroni correction \cite{bonferroni1936teoria} is a conservative post-hoc method that controls the Type I error rate, but significantly increases the probability of a Type II error. To achieve a more balanced control of Type I and Type II errors, we tested whether our results are still significant after performing the Holm-Bonferroni method~\cite{holm1979simple}. So, while we used the $p < 0.05$ level of significance to reject the null hypotheses, we also report when the significance level is still significant after applying the Holm-Bonferroni correction.

\subsubsection{Frequency Analysis and Descriptive Statistics}
\begin{figure}[htbp]
	\centering
	%trim={<left> <lower> <right> <upper>}
	{\includegraphics[width=1\linewidth]{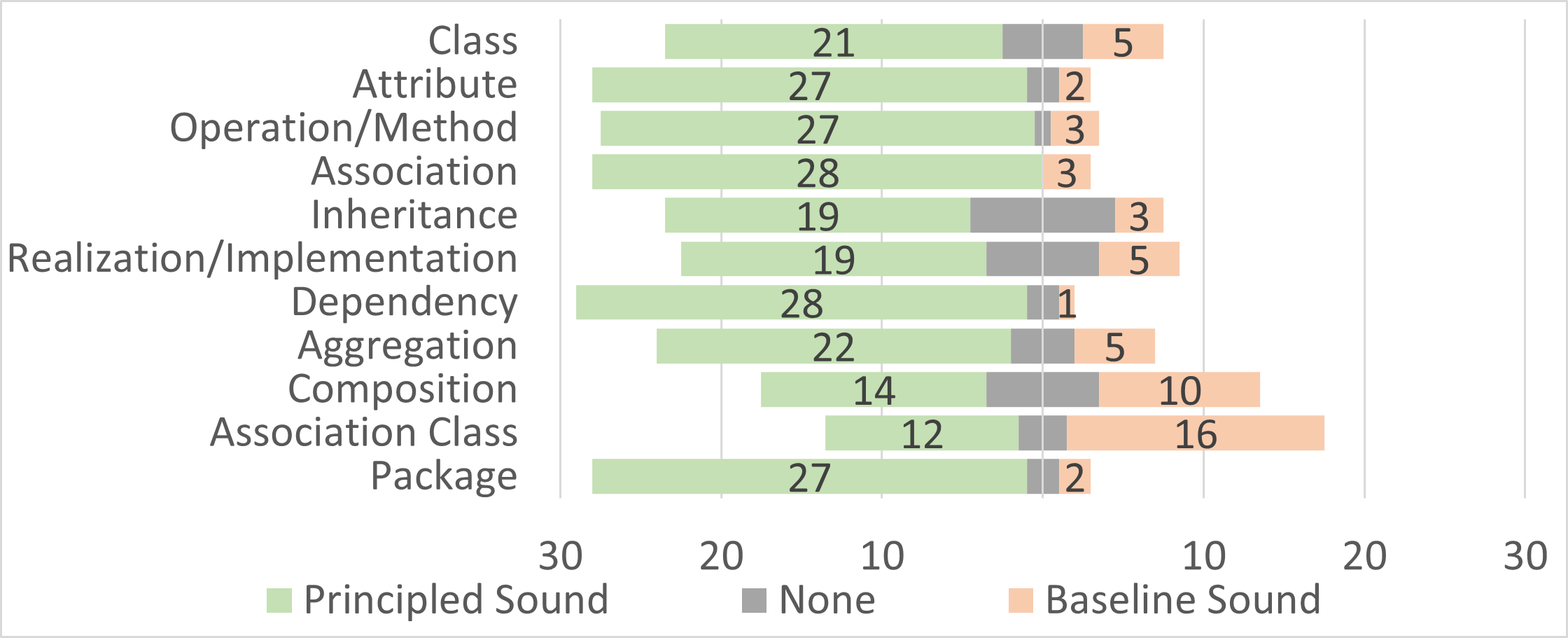}}%
	\caption{Preference expressed by participants for using the catalogue sounds vs the baseline sounds.}
	\label{fig:SoundRationaleVSRandom}
\end{figure}

%Figure \ref{fig:SoundRationaleVSRandom} presents the distribution of preferences of the participants with respect to the most adequate sound for each model element. 
Figure \ref{fig:SoundRationaleVSRandom} displays the distribution of participants' preferences for the most suitable sound for each model element.
For most model elements, there seems to be a clear preference towards the proposed sound, when contrasted to the baseline sound, or the option of suggesting a different sound. There are two notable exceptions: the Association Class sound from the baseline was chosen more often than the proposed one, and the proposed Composition element sound received a fairly balanced number of preferences, with a small advantage to the proposed sound.

\begin{figure}[htbp]
	\centering
 
%	{\includegraphics[width=1\linewidth]	
{\includegraphics[width=7.5cm]{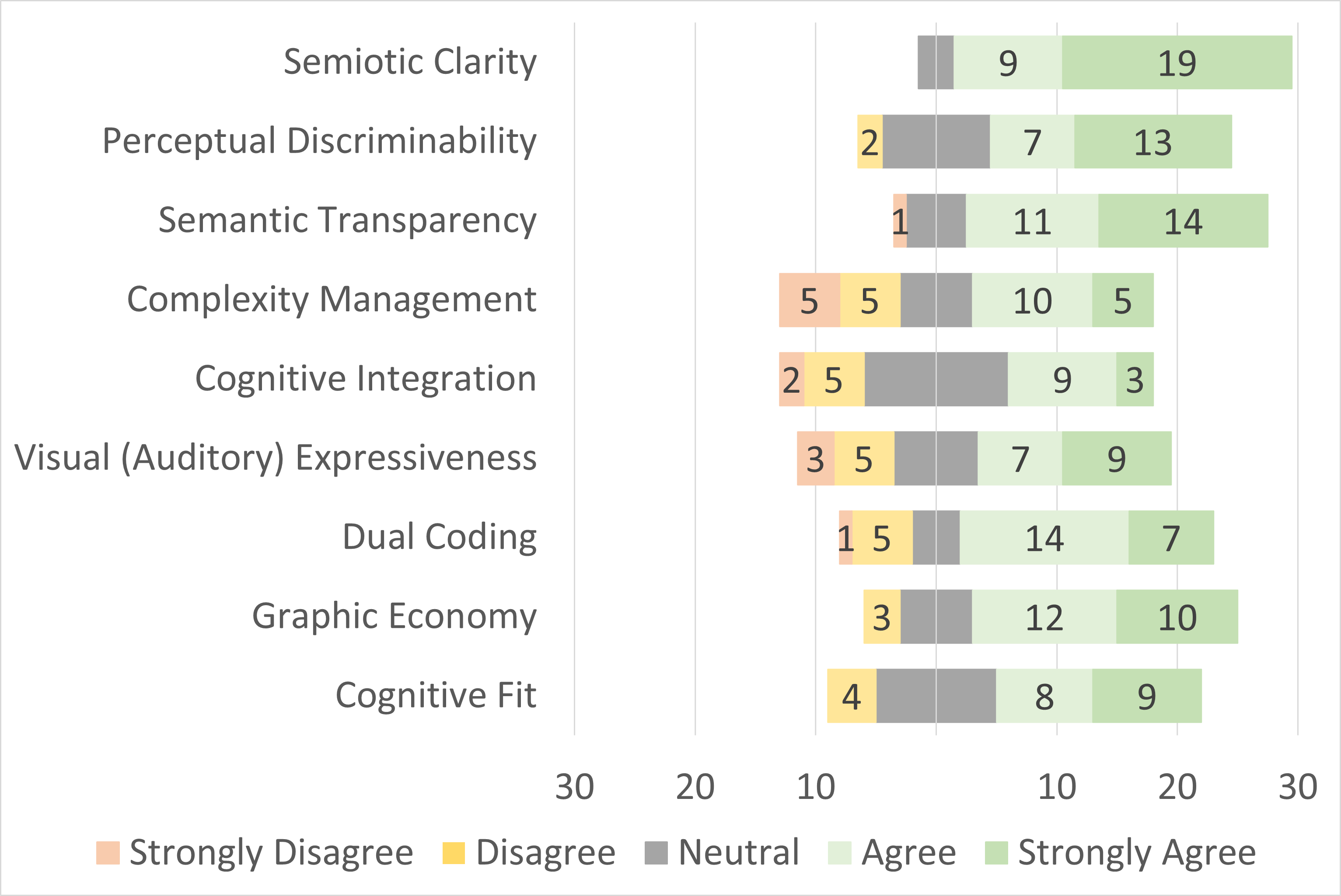}}%
	\caption{Results of the relevance of each proposed principle for Auditory notations, as rated by participants.}
	\label{fig:results_principles_ratings}
\end{figure}

The results of how participants rated the individual relevance of the proposed principles for constructing auditory notations are presented in Figure \ref{fig:results_principles_ratings}. %Overall, there are higher frequencies on the \textit{Agree} and \textit{Strongly Agree} options, suggesting an overall perception of the relevance of the proposed principles. 
 The higher frequencies in \textit{Agree} and \textit{Strongly Agree} options indicate the overall perceived relevance of the principles.
This is also observable in the descriptive statistics presented in Table \ref{tab:DescriptiveStatisticsPerceivedRelevance}, where all means are above the indifferent value of 3.

\begin{table}[htbp]
\centering
%\caption{Descriptive statistics for the perceived relevance of the principles}
\caption{Descriptive statistics for principles' perceived relevance}
\label{tab:DescriptiveStatisticsPerceivedRelevance}
\resizebox{\columnwidth}{!}{
\begin{tabular}{l|l|l|l|l|l|}
\cline{2-6}
 & \textbf{N} & \textbf{Mean} & \textbf{StdDev} & \textbf{Min} & \textbf{Max} \\ \hline
\multicolumn{1}{|l|}{\textbf{Semiotic Clarity}} & 31 & 4.5161 & .67680 & 3.00 & 5.00 \\ \hline
\multicolumn{1}{|l|}{\textbf{Perceptual Discriminability}} & 31 & 4.0000 & 1.00000 & 2.00 & 5.00 \\ \hline
\multicolumn{1}{|l|}{\textbf{Semantic Transparency}} & 31 & 4.1935 & .94585 & 1.00 & 5.00 \\ \hline
\multicolumn{1}{|l|}{\textbf{Complexity Management}} & 31 & 3.1613 & 1.34404 & 1.00 & 5.00 \\ \hline
\multicolumn{1}{|l|}{\textbf{Cognitive Integration}} & 31 & 3.1613 & 1.03591 & 1.00 & 5.00 \\ \hline
\multicolumn{1}{|l|}{\textbf{Auditory Expressiveness}} & 31 & 3.4516 & 1.33763 & 1.00 & 5.00 \\ \hline
\multicolumn{1}{|l|}{\textbf{Dual Coding}} & 31 & 3.6774 & 1.10716 & 1.00 & 5.00 \\ \hline
\multicolumn{1}{|l|}{\textbf{Graphic Economy}} & 31 & 3.9355 & .96386 & 2.00 & 5.00 \\ \hline
\multicolumn{1}{|l|}{\textbf{Cognitive Fit}} & 31 & 3.7097 & 1.03902 & 2.00 & 5.00 \\ \hline
\end{tabular}
}
\end{table}

%\vspace{-0.3cm}
\subsubsection{Hypotheses testing}
\textbf{G1.} Table \ref{tab:ChiSquareSoundPreference} presents the results of the Chi-Square goodness of fit tests concerning sound preferences. These tests were performed to determine whether the proportion of participants who preferred the sounds chosen according to the proposed auditory principles and semiotics was equal to the proportion of those preferring the baseline, or some other sound. Except for Composition, all the remaining differences are \textbf{statistically significant} even when considering the \textbf{\underline{Bonferroni-Holm correction}}, suggesting a relevant preference towards one of the sounds (the proposed sound in all but the Association Class, where our participants preferred the baseline sound). 

% \usepackage{graphicx}
\begin{table}[htbp]
\centering
\caption{Sound preference}
\label{tab:ChiSquareSoundPreference}
%\resizebox{\columnwidth}{!}{%
\begin{tabular}{l|l|l|l|}
\cline{2-4}
 & \textbf{Chi-square} & \textbf{df} & \textbf{Asymp. Sig.} \\ \hline
\multicolumn{1}{|l|}{\textbf{Class}} & 16.516 & 2 & \textbf{\underline{0.000}} \\ \hline
\multicolumn{1}{|l|}{\textbf{Attribute}} & 40.323 & 2 & \textbf{\underline{0.000}} \\ \hline
\multicolumn{1}{|l|}{\textbf{Operation / Method}} & 40.516 & 2 & \textbf{\underline{0.000}} \\ \hline
\multicolumn{1}{|l|}{\textbf{Association}} & 45.742 & 2 & \textbf{\underline{0.000}} \\ \hline
\multicolumn{1}{|l|}{\textbf{Inheritance}} & 12.645 & 2 & \textbf{\underline{0.002}} \\ \hline
\multicolumn{1}{|l|}{\textbf{Realization / Implementation}} & 11.097 & 2 & \textbf{\underline{0.004}} \\ \hline
\multicolumn{1}{|l|}{\textbf{Dependency}} & 45.355 & 2 & \textbf{\underline{0.000}} \\ \hline
\multicolumn{1}{|l|}{\textbf{Aggregation}} & 19.806 & 2 & \textbf{\underline{0.000}} \\ \hline
\multicolumn{1}{|l|}{\textbf{Composition}} & 2.387 & 2 & 0.303 \\ \hline
\multicolumn{1}{|l|}{\textbf{Association Class}} & 8.581 & 2 & \textbf{\underline{0.014}} \\ \hline
\multicolumn{1}{|l|}{\textbf{Package}} & 40.323 & 2 & \textbf{\underline{0.000}} \\ \hline
\end{tabular}%
%}
\end{table}

\textbf{G2.} Table \ref{tab:PerceivedRelevance} shows the results of the Chi-Square goodness of fit tests concerning the perceived relevance of the proposed 9 principles. These tests were performed to determine whether the proportion of participants choosing the different relevance levels is equal. 7 principles had \textbf{statistically significant differences}. Of these, 5 remain statistically significant even when considering the \textbf{\underline{Bonferroni-Holm correction}}. Overall, these results suggest, for 7 of the principles with increased confidence in 5 of them, that participants consider these principles relevant. The exceptions are Complexity Management and Auditory Expressiveness, for which opinions are more divided, even if overall there is a tendency towards agreement.

% \usepackage{graphicx}
% \usepackage[normalem]{ulem}

\begin{table}[htb]
\centering
\caption{Perceived relevance of the principles}
%\resizebox{\textwidth}{!}{%
\begin{tabular}{l|r|l|l|}
\cline{2-4}
 & \multicolumn{1}{l|}{\textbf{Chi-square}} & \textbf{df} & \textbf{Asymp. Sig.} \\ \hline
\multicolumn{1}{|l|}{\textbf{Semiotic Clarity}} & 41.742 & 4 & {\underline{\textbf{0.000}}} \\ \hline
\multicolumn{1}{|l|}{\textbf{Perceptual Discriminability}} & 17.871 & 4 & {\underline{\textbf{0.001}}} \\ \hline
\multicolumn{1}{|l|}{\textbf{Semantic Transparency}} & 24.323 & 4 & {\underline{\textbf{0.000}}} \\ \hline
\multicolumn{1}{|l|}{\textbf{Complexity Management}} & 3.032 & 4 & 0.552 \\ \hline
\multicolumn{1}{|l|}{\textbf{Cognitive Integration}} & 12.710 & 4 & \textbf{0.013} \\ \hline
\multicolumn{1}{|l|}{\textbf{Auditory Expressiveness}} & 3.355 & 4 & 0.500 \\ \hline
\multicolumn{1}{|l|}{\textbf{Dual Coding}} & 15.290 & 4 & {\underline{\textbf{0.004}}} \\ \hline
\multicolumn{1}{|l|}{\textbf{Graphic Economy}} & 15.613 & 4 & {\underline{\textbf{0.004}}}\\ \hline
\multicolumn{1}{|l|}{\textbf{Cognitive Fit}} & 11.097 & 4 & \textbf{0.025} \\ \hline
\end{tabular}%
%}
\label{tab:PerceivedRelevance}
\end{table}

\subsection{Discussion Of Results}

\subsubsection{Participants' preferences of sound catalogue (G1)}
%In this subsection, the observed results related to the catalogues and the choices of sounds made by the participants for the different diagrammatic elements that constitute UML class diagrams, are discussed.

%According to the results obtained and summarised in the table \ref{tab:results_table}, we can verify that in a total of 11 questions asked, where each corresponds to a different element of the UML class diagrams, the sounds that were proposed according to the catalogue of good practices was the choice with the highest number of votes in 90.91\% of the questions, that is, in 10 of them. In 9 of these 10 questions, the result was above 50\%, thus obtaining an absolute majority, while in 1, the registered figure was 45.16\%. The sounds for the catalogue with unsatisfied auditory principles were the option with the most votes in 9.09\% of all questions. This information can be found in table \ref{tab:chosen_sounds_catalogue}.

The sounds proposed in the catalogue of good practices were chosen by the majority 10 out of 11 times (90.91\%). Among these, 9 questions had an absolute majority of over 50\%, while 1 had 45.16\%. The baseline catalogue with unsatisfied auditory principles received the most votes in only one question. %See table \ref{tab:chosen_sounds_catalogue} for more details.

In 7 out of the 11 questions (63.64\%) the sounds proposed in the catalogue of good practices received a high preference rate of over 65\% from the participants. These sounds corresponded to elements such as \textit{class}, \textit{attribute}, \textit{operation/method}, \textit{association}, \textit{dependency}, \textit{aggregation}, and \textit{package}. These results suggest that these sounds are more suitable to represent the diagrammatic elements in question compared to the sounds in the catalogue of unsatisfied principles. Participants intuitively grasped the rationale for selecting those sounds, and the number of answers for the unsatisfied auditory principles and \textit{None} options was minimal. Only two sounds violating the auditory principles were chosen by more than 5 participants (16.13\%). The number of answers for the \textit{None} option was relatively low, but participants did offer some suggestions and alternatives for sounds for the model elements. For example, two participants suggested the sound of a baby for \textit{inheritance}. While the rationale is clear, this would violate the principle of Semiotic clarity, as there is already a sound of a baby crying used in the concept dependency (which was among the most preferred sounds). Another participant suggested the usage of an orchestra sound for the element of \textit{composition}. Again, the rationale is clear and this could be considered for an evolution of our proposed catalogue, as it does not clash with other sounds.

Some participants who chose \textit{None} as an answer did not suggest any alternative or had derivative answers, which would violate the principle of semiotic clarity. Some participants suggested only minor modifications, which may be due to their lack of expertise in suggesting new sounds as computer engineers. Involving individuals with a musical or sound engineering background could improve results in the future.

The least expressive figures in the sound catalogue that followed semiotics and proposed auditory principles were found in questions \textit{composition} and \textit{association classes}, with values of 45.16\% and 38.71\%, respectively. In the latter, most participants chose the sound used for the unsatisfied principles catalogue, suggesting that the sound we proposed is not suitable and should be replaced in a future version of the catalogue. It may also be the case that the element \textit{association class} is less well understood by participants, which would make it harder to associate it with any given sound. 

%Conclusão
%Overall, we consider that the results obtained were encouraging, since for 90.91\% of the sounds, the subjects considered the ones proposed based on the auditory principles and semiotics to be more adequate when compared to the sounds used in the catalogue of unsatisfied principles, thus fulfilling the objective of being more adequate choices to represent in an auditory manner, the diagrammatic elements in question. Despite this, it should be taken into consideration that certain choices of sounds may be even better than those proposed in this study since even though all the sounds here considered try to accommodate the universal human experience in order to be as understandable as possible, different people have different sensibilities and different life experiences, so there may always be certain sounds that won't be the most adequate for everyone. However, the main objective of this study was to observe whether there were indeed differences when using a more informed method for the construction of sounds for auditory notations in contrast to a naive approach. In that respect, the results here obtained suggest that there are indeed positive and noticeable differences. However, we think that this study would benefit from a larger sample size so that these questions could be further explored.

%The results obtained were encouraging, with 90.91\% of the sounds proposed based on auditory principles and semiotics being considered more adequate than the sounds in the catalogue of unsatisfied principles. 

Our goal was to compare the benefits of using an informed sound selection method to a naive approach, and the results suggest positive and noticeable differences. It is safe to assume the proposed catalogue can be iteratively improved using A/B testing~\cite{siroker2015b}, in the future, but we can now leverage a set of principles to propose more ``promising'' alternatives.

%-----Principles
\subsubsection{Relevance Of The Principles For Constructing Auditory Notations (G2)}
Most of the auditory principles were perceived as at least somewhat relevant by our participants. The two exceptions were \textit{Complexity Management} and \textit{Auditory Expressiveness}.
The top three most relevant principles were \textit{Semiotic Clarity}, \textit{Semantic Transparency}, and \textit{Perceptual Discriminability}. These principles received the strongest emphasis in the sound choices and were frequently unsatisfied in the catalogue of bad practices. Results suggest that participants' exposure to a principle correlates with how relevant they consider it. Principles that could not be tested in the study, such as Cognitive Integration and Complexity Management, were considered less relevant. Overall, the proposed auditory principles were well received, with 93.5\% of participants having a positive view of their relevance, supporting the hypothesis that they are relevant from the user's perspective.

\subsection{Threats To Validity}

Although the proposed sounds in the catalogue that satisfy auditory principles attempt to accommodate the universal human experience for better comprehension, individual sensitivities, cultural backgrounds, and life experiences may make some sounds less suitable for certain individuals. The selection of sounds in the catalogue of unsatisfied principles may also be biased, as the intention was to choose sounds that do not follow the proposed principles. Additionally, limitations in the study design, such as the use of a web-based platform for sound playback and potential distractions for participants (the surrounding noise was not controlled), may affect the results. Testing all the proposed principles was not feasible, and convenience sampling may have excluded certain profiles.

\section{Conclusions and Future work}
\label{sec:conclusion}

%After analyzing previous studies on sound design, we adapted Moody's work on  \textit{The "Physics" Of Notations} from visual to auditory principles. This produced a set of guidelines and principles for creating auditory notations. Using this as a foundation, we proposed a catalogue of UML Class Diagrams sounds.
%We conducted an experiment comparing the proposed catalogue of designed sounds according to the guidelines to an arbitrary catalogue that ignored auditory principles and semiotics. A user study was also conducted to gauge the relevance of the proposed principles. Overall, the results were encouraging, with respondents finding the sounds based on auditory principles and semiotics to be more suitable. This suggests that using an informed approach to sound design for auditory notations is beneficial. Participants also viewed the principles as relevant.

We adapted Moody's \textit{The "Physics" Of Notations} from visual to auditory principles, producing a set of guidelines for creating auditory notations. Based on these guidelines, we proposed a catalogue of UML Class Diagrams sounds and conducted an experiment comparing it to a baseline catalogue that ignored auditory principles and semiotics. Results showed that sounds based on auditory principles and semiotics were found to be more suitable, with participants viewing the principles as relevant. This suggests that an informed approach to sound design for auditory notations is beneficial.

With the growing complexity of software systems, effective communication between designers and stakeholders has become more important than ever. The use of sound-based notations can support an intuitive multimodal, and efficient way to convey complex information, making it easier for non-experts to understand and provide feedback. This study represents an important first step towards the development of more effective sound-based notations. It provides a foundation for further research in this area that could ultimately have a significant impact on modelling tools.

\section{Acknowledgements}

This work is supported by NOVA LINCS (UIDB/04516/2020) with the financial support of FCT.IP

\bibliographystyle{./IEEEtran}
\bibliography{./bibliography.bib}

\end{document}